# Stabilizing the Tb-based 214 cuprate by partial Pd substitution

Yuzki M. Oey[1], James Eujin Park[1], Jing Tao[2], Elizabeth M. Carnicom[1], Tai Kong[1], Marisa B. Sanders[1], and R. J. Cava[1]

[1]*Department of Chemistry, Princeton University, Princeton, NJ 08540*

[2]*Condensed Matter Physics and Materials Science Division, Brookhaven National Laboratory, Upton, NY 11973*

*Abstract*

Previously known to form only under high pressure synthetic conditions, here we report that the T'-type 214-structure cuprate based on the rare earth atom Tb is stabilized for ambient pressure synthesis through partial substitution of Pd for Cu. The new material is obtained in purest form for mixtures of nominal composition $Tb_{1.96}Cu_{0.8}Pd_{0.2}O_4$. The refined formula, in orthorhombic space group *Pbca*, with $a = 5.5117(1)$ Å, $b = 5.5088(1)$ Å, and $c = 11.8818(1)$ Å, is $Tb_2Cu_{0.83}Pd_{0.17}O_4$. An incommensurate structural modulation is seen along the *a* axis by electron diffraction and high resolution imaging. Magnetic susceptibility measurements reveal long range antiferromagnetic ordering at 7.9 K, with a less pronounced feature at 95 K; a magnetic moment reorientation transition is observed to onset at a field of approximately 1.1 Tesla at 3 K. The material is an n-type semiconductor.

## Introduction

The $R_2CuO_4$ ("214") rare earth (R) cuprates have been of interest due to their remarkable properties.[1,2] Long-range magnetic ordering of the rare earth moments has been observed for a variety of these materials, including $R$ = Gd, Tb, Dy, Ho, Er, and Tm.[3–5] Most importantly, with the appropriate partial substitutions, $R_2CuO_4$ compounds with $R$ = La, Pr, Nd, and Sm become *n*- or *p*- type high $T_C$ superconductors.[6,7] $R_2CuO_4$ compounds based on the rare earths smaller than Gd have previously been synthesized only at high pressures, with at least 8 GPa applied pressure required to form 214 cuprates for $R$ = Tb, Dy, Ho, and Er; no $R_2CuO_4$-formula compounds are known to be stable at ambient pressure for the small rare earths.[6,8,9] The materials display a rich evolution of structure based on the size of the rare earth ion. The structures range from the so-called T-type structure of $La_2CuO_4$, in which the Cu is in distorted octahedral coordination, through the T*-type structure typified by $Nd_{2-x-z}Ce_xSr_zCuO_4$ in which the Cu is in pyramidal coordination, through the T' structure displayed by $Nd_2CuO_4$ in which the Cu is in square planar coordination.

The T'-type 214 structure, which is of interest here, is most often reported as body centered tetragonal, with $a_0 \sim 3.9$ Å and $c_0 \sim 11.0$ Å, although there are reports of orthorhombic symmetry for T'-type $Gd_2CuO_4$ in a side-centered space group with $a' \sim b' \sim \sqrt{2} \times a_0$, and $c' \sim c_0$.[10,11] Three supercells have been reported for the T' structure, with the most common (A) supercell having cell parameters of $2\sqrt{2}a_0 \times \sqrt{2}a_0 \times c_0$.[9] The current study is based on the hypothesis that some of the rare-earth-based 214 cuprates previously formed only at high pressures might form at ambient pressure through a partial substitution of $Pd^{2+}$ for some of the $Cu^{2+}$ present. $Pd^{2+}$ is a potential stabilizing ion because its strong Jahn-Teller ($4d^8$) character causes it to strongly prefer square planar coordination with oxygen, a characteristic of Cu in the T' structure.

Consistent with this hypothesis, here we report the synthesis and characterization of the Tb-based T' 214 phase stabilized for preparation at ambient pressure via 20% Pd substitution for Cu. This phase could not be prepared at ambient pressure with no Pd present. Variability of a few percent in the Cu to Pd ratio may be allowed in the phase, but we did not study that in detail in the current work. We were able to prepare the compound most cleanly as a bulk material for mixtures that are 2% Tb deficient, but the refined Tb content of the 214 phase is stoichiometric, as is the oxygen content. Our Rietveld refinement of ambient temperature synchrotron powder X-ray diffraction data identified the phase to form in the orthorhombic *Pbca* space group, with cell parameters $a' \sim b' \sim \sqrt{2} \times a_0$, and $c' \sim c_0$. An incommensurate superstructure reflecting the presence of a structural modulation of approximately twice the *a* axis was seen in ambient temperature electron diffraction measurements. The magnetic data is generally similar to that of other T' structure cuprates.[3]

**Experiment**

The samples were synthesized using the starting materials $Gd_2O_3$ (Anderson Physics Laboratories, 99.99%), $Tb_4O_7$ (Alfa Aesar, 99.9%), $Dy_2O_3$ (Alfa Aesar, 99.99%), PdO (Aldrich Chemical Company, Inc., 98%), and CuO (Alfa Aesar, 99.7%). The rare earth oxides were dried at 800 °C in air for at least 24 hours before use. The $Tb_4O_7$ was reduced to $Tb_2O_3$ before use at 700 °C under 5% $H_2$/Ar (Airgas) for 24 hours to ensure initial homogeneity in oxidation state and more precise starting material compositions. Selected quantities of the starting materials (precision of weighing was ± 0.0005 g) were ground and mixed in 0.5 g batches with an agate mortar and pestle, then heated in an alumina crucible at 1000 °C under air for 36 hours, then at

1050 °C under air for 36 hours with intermediate grinding. All products are stable in air up to 1050 °C.

The purity and structures of the samples were examined at room temperature by laboratory powder X-ray diffraction (pXRD) using a Bruker D8 Advance Eco diffractometer (Cu Kα radiation, λ = 1.5406 Å) with a LynxEye-XE detector. A Quantum Design Physical Property Measurement System (PPMS) Dynacool equipped with a vibrating sample magnetometer (VSM) option was used to measure temperature-dependent (1.7–300 K, H = 1000 Oe and 50 Oe) and field-dependent ($\mu_0 H$ = 0 to 9 T at 3, 50, and 200 K) magnetization. Susceptibility is defined as M/H. The powder sample was packed into a plastic sample holder and loaded onto a brass sample rod for the susceptibility measurements. The PPMS was also used for temperature-dependent (250–300 K) four probe resistivity measurements with a constant current of 0.050 mA. A pellet of the sample was pressed at 1 metric ton/0.317 $cm^2$ area using a pellet die and a hydraulic press, and densified at the synthesis condition. The pellet was then cut into rectangles and Pt wires were attached using silver paint. The carrier type was determined at room temperature by a hot-probe Seebeck coefficient test. In this test, the positive and negative terminals of an ammeter were attached to opposite sides of a pellet and a preheated soldering iron was touched to the pellet near the negative terminal. To measure the oxygen content, a SDT Q600 DSC/TGA was used on a powder sample heated at 0.25 C per minute in a flowing 10%$H_2$ 90%Ar gas mixture. Synchrotron X-ray diffraction data was taken on a powder sample at the Advanced Photon Source (APS) on beamline 11-BM with a wavelength λ = 0.412667 Å at 295 K. To determine the atomic structural parameters, a Rietveld refinement of the synchrotron data was performed using the FULLPROF diffraction suite with Thompson-Cox-Hastings pseudo-Voigt peak shapes. Electron diffraction experiments were carried out at Brookhaven National

Laboratory using a double-Cs corrected JEOL ARM 200F transmission electron microscope (TEM). TEM samples of $Tb_2Cu_{0.83}Pd_{0.17}O_4$ were prepared directly from the bulk material by crushing and grinding for electron transparency. Energy dispersive X-ray spectroscopy (EDX) was employed to examine the elemental distributions in different particles in the TEM samples. The EDX results (not shown) indicate chemical homogeneity and a consistent stoichiometry of the examined fragments.

**Results and Discussion**

Representative room temperature laboratory powder X-ray diffraction (pXRD) data, employed to identify and then determine the approximate formula of the new T' material, are shown in Figures 1, 2, and 3. Initially, $Tb_2O_3$ and CuO were combined in a 2:1 Tb to Cu ratio and heated in air. The resulting pXRD pattern primarily shows $Tb_2Cu_2O_5$ (Fig. 1a), with no T' phase present. However, when CuO is partially substituted by PdO in the starting mixture, a new phase of the T' 214 type appears in the resulting pXRD patterns. With a substitution of $x = 0.05$, the XRD pattern includes the signature of a T'-like 214 phase with the main peak at 32.2° (indicated by the asterisk), but significant amounts of $Tb_{11}O_{20}$ and $Tb_2Cu_2O_5$ are present as well (Fig. 1b). When Pd is substituted at a higher ratio, $x = 0.20$, the new phase becomes dominant with minimal $Tb_{11}O_{20}$ impurity (peak at 29.1°, indicated by the black triangle (Fig. 1c)). However, increasing the Pd substitution to $x = 0.35$ (Fig. 1d) results in increasing the fraction of impurity. Thus the Tb-based 214 phase, previously reported as only forming at high pressures, is formed at ambient pressure when stabilized by partial Pd-substitution for Cu, consistent with our initial hypothesis.

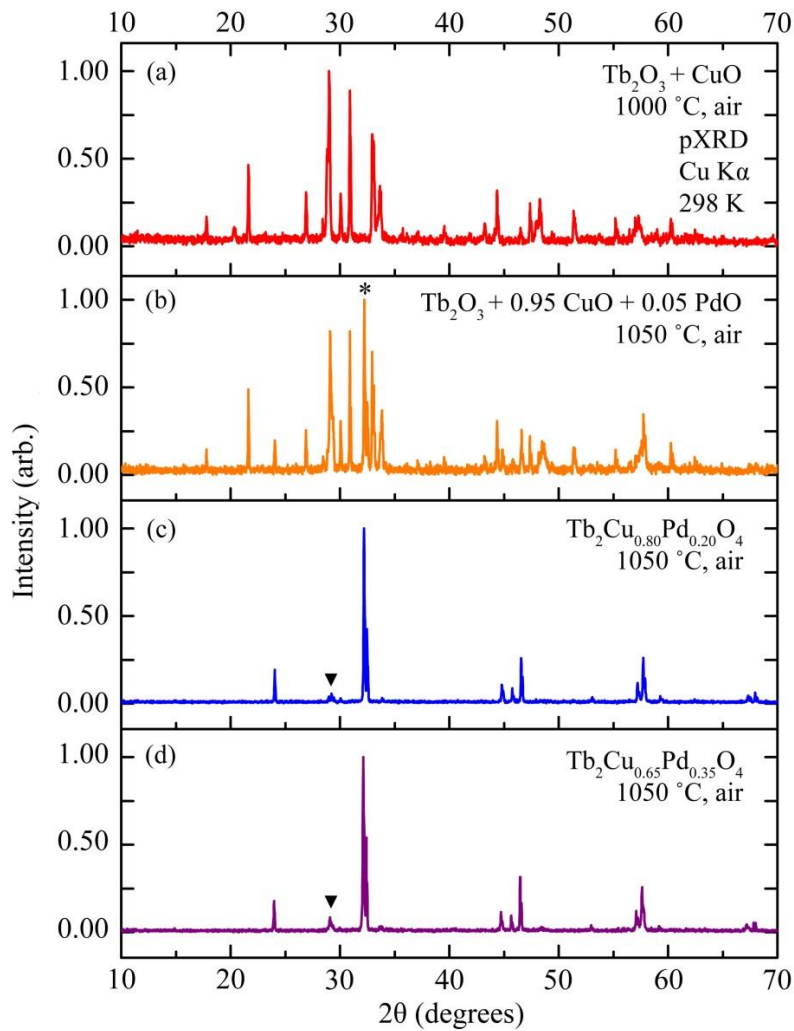

FIG. 1. Room temperature powder XRD patterns for $Tb_2Cu_{1-x}Pd_xO_4$ ($x$ = 0, 0.05, 0.20, and 0.35) synthesized at ambient pressure. (a) With $x$ = 0, only $Tb_2Cu_2O_5$ and other phases are formed. (b) A Pd doping level of $x$ = 0.05 shows a new peak (*) at $2\theta$ = 32.2° due to the T' phase but a significant amount of $Tb_2Cu_2O_5$ as well. (c) At $x$ = 0.20, the T' 214 phase is dominant (the unmarked diffraction peaks) with a small $Tb_{11}O_{20}$ impurity (▼). (d) However, with $x$ = 0.35, $Tb_{11}O_{20}$ (▼) grows with respect to the desired T' phase. In (c) and (d), all unmarked peaks are due to the T' phase.

We also attempted to synthesize $Gd_2Cu_{0.8}Pd_{0.2}O_4$ and $Dy_2Cu_{0.8}Pd_{0.2}O_4$, materials based on the rare earths directly to the left (larger) and right (smaller) of Tb in the rare earth series by the same method. The pXRD pattern of $Gd_2Cu_{0.8}Pd_{0.2}O_4$ in Fig. 2a agrees with the published pattern for $Gd_2CuO_4$ prepared at ambient pressure, with a slight downshift of the peaks in the diffraction pattern due to an increase in lattice size accompanying the incorporation of Pd on the Cu site. The peaks for $Tb_2Cu_{0.8}Pd_{0.2}O_4$ (Fig. 2b) align nicely with those of $Gd_2Cu_{0.8}Pd_{0.2}O_4$, showing the successful formation of a Tb-based 214 phase. $Dy_2Cu_{0.8}Pd_{0.2}O_4$, on the other hand, could not be prepared under the same synthesis conditions (Fig. 2c). Instead, $Dy_2O_3$ and $Dy_2Cu_2O_5$ were formed, indicating that partial Pd substitution could not stabilize a T' phase in this chemical system at ambient pressure. Thus the stabilization of the 214 phase for ambient pressure synthesis through partial Pd substitution is only possible for the rare earth Tb, which is directly at the border between the 214 cuprates that are synthetically accessible at ambient pressures and those synthetically accessible at high pressures.

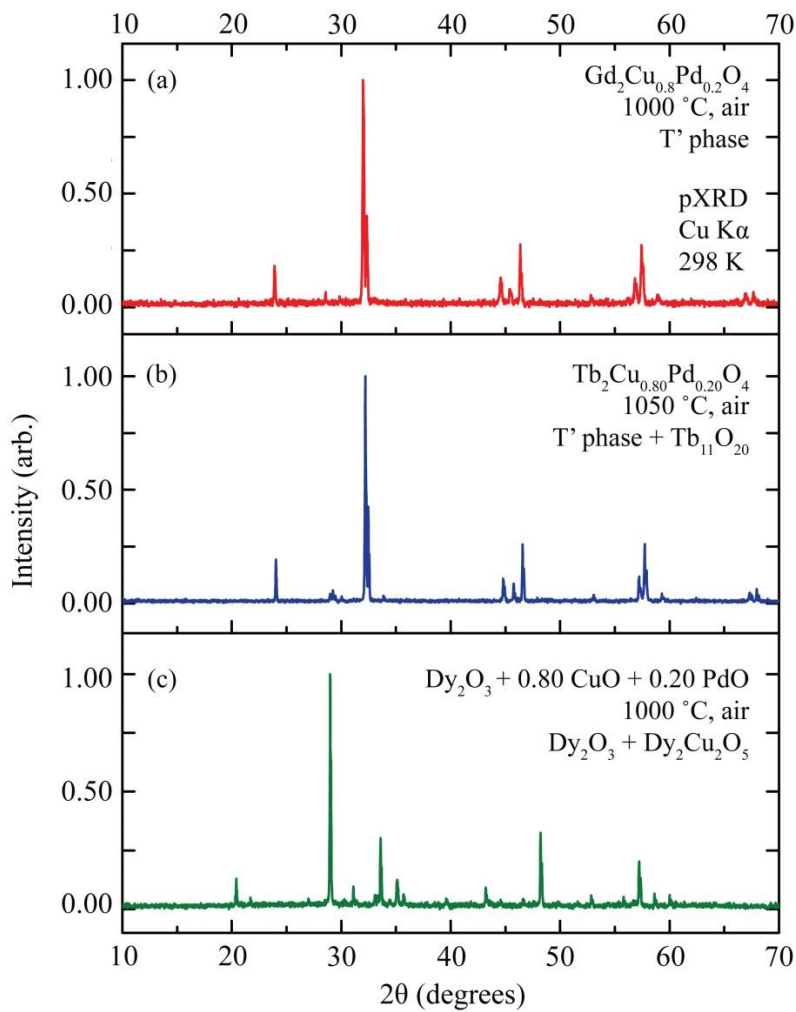

FIG. 2. Room temperature powder XRD patterns for different partially Pd-substituted 214 ratio rare earth cuprate materials prepared under the current conditions. (a) The T' peaks are shifted to lower angles from the characteristic peak of $Gd_2CuO_4$ due to the partial Pd substitution. (b) An almost identical set of peaks is seen for $Tb_2Cu_{0.8}Pd_{0.2}O_4$, at slightly higher angles due to the smaller size of Tb compared to Gd, with a small impurity of $Tb_{11}O_{20}$. (c) A mixture of $Dy_2O_3$, CuO, and PdO did not yield the 214 T' type phase, instead primarily forming $Dy_2O_3$ plus $Dy_2Cu_2O_5$.

The purity of bulk T' phase samples was further improved by varying the Tb content in the starting mixture. The pXRD patterns of $Tb_yCu_{0.8}Pd_{0.2}O_4$ samples with $y$ = 2, 1.96, and 1.92 are shown in Fig. 3. With $y$ = 2 (Fig. 3a), some $Tb_{11}O_{20}$ impurity is present, evidenced by the diffraction peak at 29.1° (indicated by the black triangle). When the Tb content was decreased from $y$ = 2 to 1.96 (Fig. 3b), this impurity peak decreased. However, when the Tb content was further reduced to $y$ = 1.92 (Fig. 3c), the impurity peak reappeared. Thus, the optimum nominal composition needed to synthesize the purest bulk sample of the Tb-based T' phase was determined to be $Tb_{1.96}Cu_{0.8}Pd_{0.2}O_4$. We are aware that this is a small deviation from the ideal rare earth stoichiometry, but to the best of our experimental ability this represents the best nominal composition for preparing a bulk sample of the T'-type 214 phase in this system under our conditions. We note that rare earth deficiency has been reported for the Nd-based 214 cuprate.[12] Our synchrotron-diffraction-based crystal structure refinements, described below, found the actual formula of the T' phase to be $Tb_2Cu_{0.83}Pd_{0.17}O_4$ so we employ that composition to describe the compound in the remainder of this report.

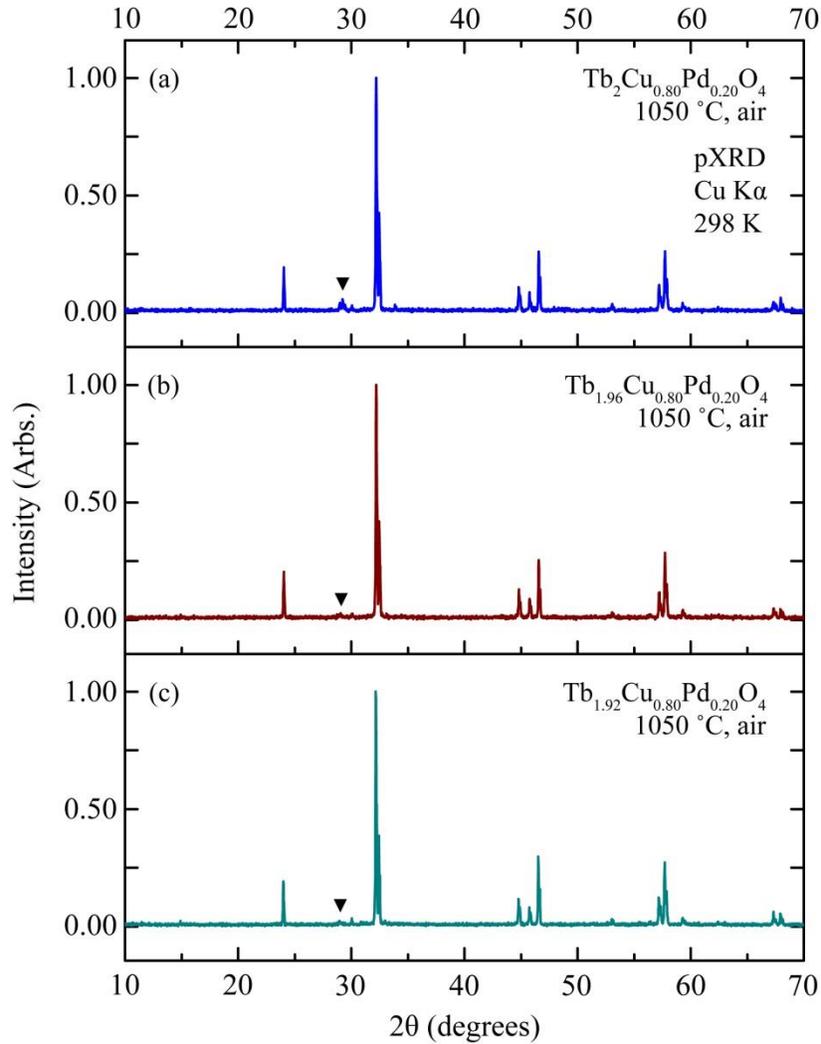

FIG. 3. Powder XRD patterns taken at room temperature for $Tb_yCu_{0.8}Pd_{0.2}O_4$ samples synthesized at ambient pressure at 1050 °C in air. All unmarked peaks are from the 214-type T' phase, with the strongest peak at $2\theta = 32.2°$. The impurity $Tb_{11}O_{20}$ phase is marked with an inverted triangle (▼) at $2\theta = 29.1°$. (a) With $y = 2$, the impurity peak is clearly present. (b) When the nominal Tb content is reduced to $y = 1.96$, the impurity almost disappears, although it is still present. (c) However, with $y = 1.92$, the impurity peak grows again with respect to the desired T' phase.

Finally, we performed thermogravimetric analysis of $Tb_2Cu_{0.83}Pd_{0.17}O_4$ heated under 5% $H_2$/Ar gas (Fig. 4) to determine the oxygen content. The compound was determined to reduce to $Tb_2O_3$, Cu metal, and Pd metal after treatment by powder X-ray diffraction. From the mass loss, the oxygen content was determined to be 3.99 O per formula unit. With an estimated error of ± 0.01 O per formula unit, this value is not significantly different from the ideal 214 T' phase oxygen content of 4.00.

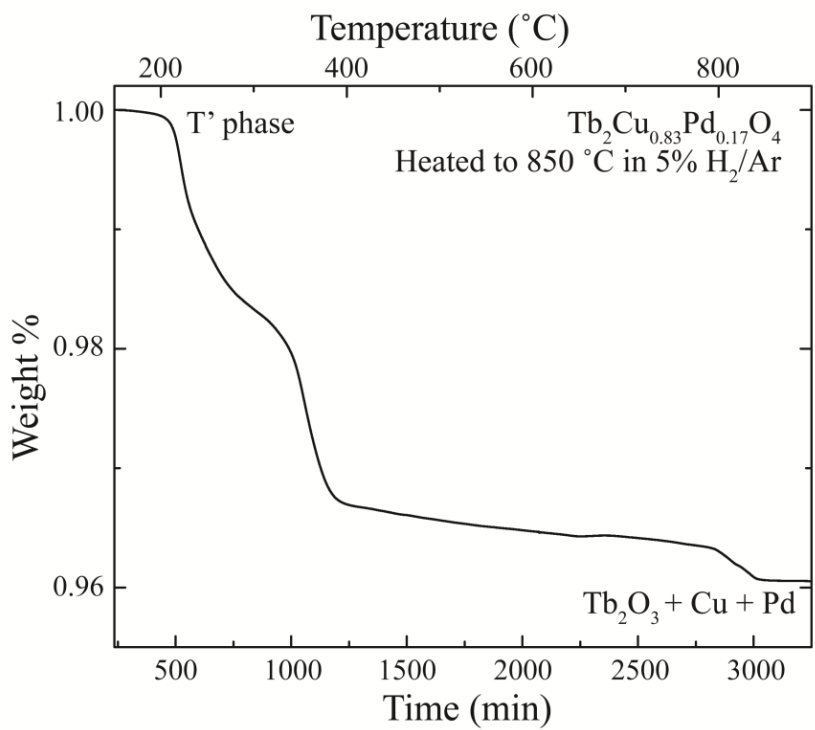

FIG. 4. Thermogravimetric analysis of a powder sample of $Tb_2Cu_{0.83}Pd_{0.17}O_4$ reduced under flowing 5% $H_2$/Ar from 150–850 °C at 0.25 °C/min. To determine the oxygen content, the final products were assumed to be $Tb_2O_3$, Cu, and Pd, with all mass loss attributed to oxygen loss.

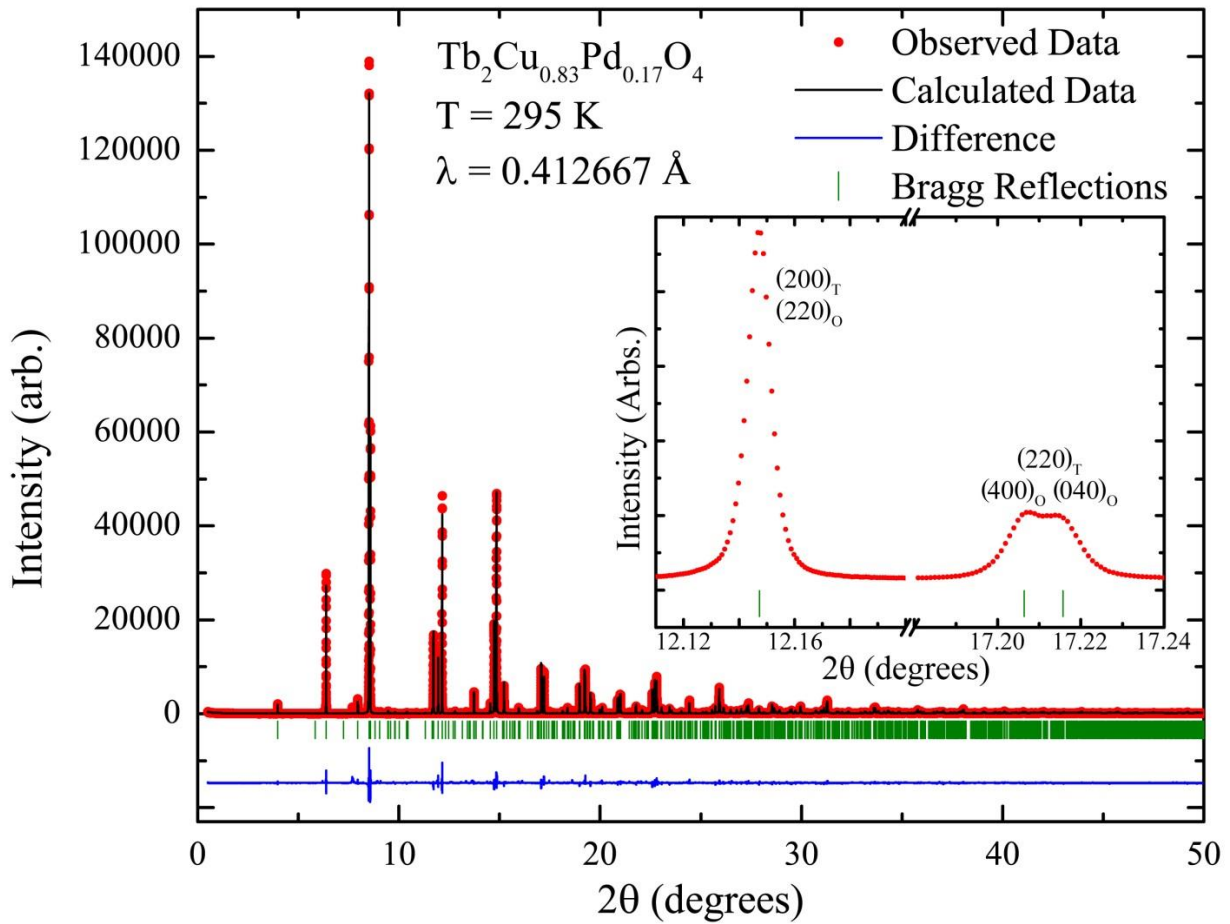

FIG. 5. Rietveld refined (*Pbca*) 11-BM synchrotron powder diffraction data of $Tb_2Cu_{0.83}Pd_{0.17}O_4$ taken at 295 K. Observed data are shown as circles, Rietveld profile as the solid line, their difference at the very bottom, and the calculated Bragg reflections as tick marks. The inset displays a split peak between the (400) and (040) peaks at 17.21° which indicates that the material is orthorhombic (O, *Pbca*) not tetragonal (T, *I4/mmm*). The singular (220) ortho (O) peak and the split (400) (040) ortho peaks indicate that the unit cell axes are rotated with respect to the tetragonal cell axes such that $a' \sim b' \sim \sqrt{2} \times a_0$.

To more qualitatively describe the structure of the Pd stabilized T'-structure Tb cuprate, high-resolution-high-intensity synchrotron diffraction data for $Tb_2Cu_{0.83}Pd_{0.17}O_4$ was fit using Rietveld structure refinements (Fig. 5). The symmetry of the material is not tetragonal, evidenced by the peak splitting of some of the normally single reflections for a tetragonal symmetry structure (inset of Fig. 5). Moreover, although the tetragonal (200) peak does not split, the tetragonal (220) peak does (Fig. 5 inset), indicating a 45° rotation of axes such that $a'$ and $b'$ ~ $\sqrt{2} \times a_0$ but are not equal to each other, identifying the structure as orthorhombic. When the high symmetry orthorhombic space group *Fmmm* (No. 69) was initially used to model the structure, based on transforming the positions in the tetragonal T' cell into the larger, rotated orthorhombic cell, a reasonable fit to the pXRD data was obtained, but the atom positions do not differ from those allowed in *I4/mmm* and thus did not provide an atom-level justification for the orthorhombic structure. Due to the report of rotations of the $CuO_4$ squares in the Gd-based 214 cuprate, refinements using space group *Cmce* (No. 64) were also run, generating a better fit than *Fmmm*.[11] However, an additional decrease of symmetry to *Pbca* (No. 61) further improved the fit and thus yielded the best structural model. Based on the refinements in this space group, the oxygen positions are significantly shifted off the ideal $x = 1/4$ $y = 1/4$ sites of the higher symmetry *Fmmm* structure, with the new sites found to be (0.210(1) 0.284(1) 0.0112(5)) and (0.262(1) 0.255(4) 0.2394(4)) as opposed to (1/4 1/4 0) and (1/4 1/4 1/4). This is primarily due to the rotation of the $Cu/Pd-O_4$ squares as described further below. Free refinement of the Tb site occupancy did not yield any deviation from full occupancy and thus the T' phase is determined to be stoichiometric in Tb content. Similarly, in free refinements, 17% of the Cu sites were found to be substituted by Pd. This number is distinguished as beyond statistical error from the nominal

20% value, thus establishing the formula of the T' material as $Tb_2Cu_{0.83}Pd_{0.17}O_4$. The final structural parameters are shown in Table I.

TABLE I. Average structure of $Tb_2Cu_{0.83}Pd_{0.17}O_4$ at 295 K with $a$ = 5.5117(1) Å, $b$ = 5.5088(1) Å, $c$ = 11.8818(1) Å, and V = 360.7626(9) Å$^3$.

| | *Fmmm* (No. 69) refined parameters (initial structural model) | | | | | |
|---|---|---|---|---|---|---|
| Atom | Wyckoff Position | $x$ | $y$ | $z$ | $B_{iso}$* | Occupancy |
| Tb1 | 8$i$ | 0 | 0 | 0.34706(3) | 0.497(3) | 1 |
| Cu | 4$a$ | 0 | 0 | 0 | 0.107(1) | 0.82(2) |
| Pd | 4$a$ | 0 | 0 | 0 | 0.107(1) | 0.18(2) |
| O1 | 8$f$ | 1/4 | 1/4 | 1/4 | 1.231(6) | 1 |
| O2 | 8$e$ | 1/4 | 1/4 | 0 | 1.231(6) | 1 |

| | *Pbca* (No. 61) refined parameters (final structural model) | | | | | |
|---|---|---|---|---|---|---|
| Atom | Wyckoff Position | $x$ | $y$ | $z$ | $B_{iso}$* | Occupancy |
| Tb1 | 8$c$ | 0.0026(2) | -0.0007(4) | 0.34702(2) | 0.504(5) | 1 |
| Cu | 4$a$ | 0 | 0 | 0 | -0.07(2) | 0.416(3) |
| Pd | 4$a$ | 0 | 0 | 0 | -0.07(2) | 0.084(3) |
| O1 | 8$c$ | 0.262(1) | 0.255(4) | 0.2394(4) | -0.44(5) | 1 |
| O2 | 8$c$ | 0.210(1) | 0.284(1) | 0.0112(5) | -0.44(5) | 1 |

*The thermal parameters of Cu and Pd are constrained to be equal in the refinement, as are the thermal parameters of O1 and O2.

The crystal structure based on the results of the *Pbca* refinement (Fig. 6) shows the presence of a very small dimensional distortion of the Cu/Pd-O$_4$ squares, with bond lengths of 1.95(1) and

2.00(1) Å. Further, there is a small but significant puckering of the squares. The primary feature of interest is the rotation of the Cu/Pd-$O_4$ squares. This kind of rotation is ubiquitous in perovskites, and in the current material is due to the size mismatch between the Tb-O and Cu/Pd-O polyhedra. Very small peaks, less than 1% of the primary peak intensity, are present in some regions of the synchrotron powder diffraction pattern; they do not correspond to any known phases in this chemical system, and were not indexed by an orthorhombic space group with no systematic absences or by hypothesizing the presence of an (A) type supercell. We tentatively attribute them to the incommensurate superlattice observed by high resolution transmission electron microscopy (see below), to which we also tentatively attribute the origin of the anomalous thermal displacement parameters for the oxygen atoms in the final structural model.

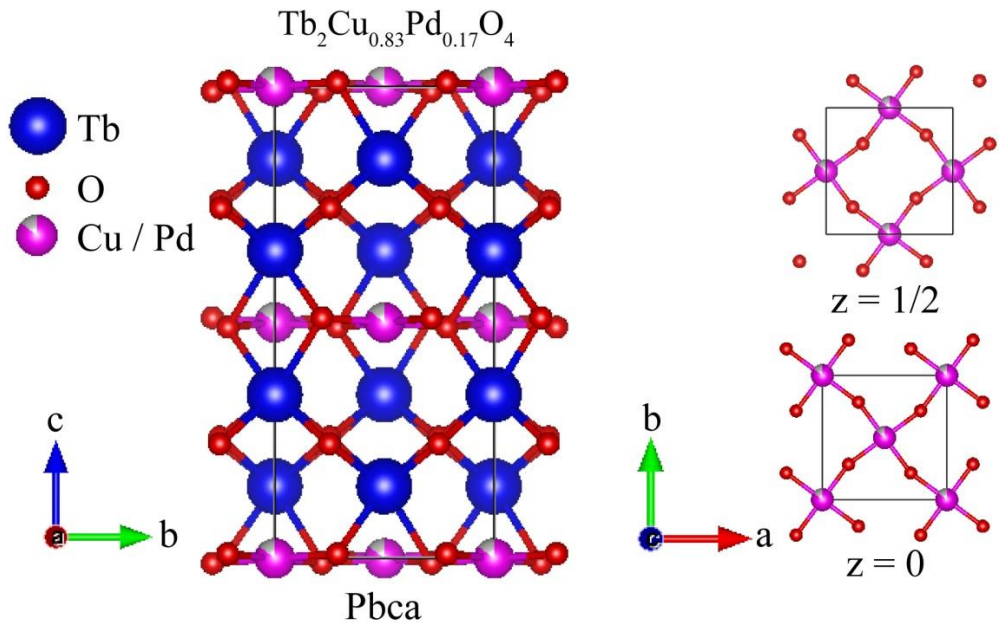

FIG. 6. (Left) The orthorhombic crystal structure for $Tb_2Cu_{0.83}Pd_{0.17}O_4$ based on the results of the refinements in space group *Pbca*. (Right) The Cu/Pd-$O_4$ squares are rotated in opposite senses in the $z = 0$ and $z = 1/2$ planes. The rotation of the $MO_4$ squares is not found in 214 structures reported in the *I4/mmm* space group.

Electron microscopy results obtained from a crushed $Tb_2Cu_{0.83}Pd_{0.17}O_4$ sample are shown in Fig. 7. Fig. 7a shows the electron diffraction pattern from a particle oriented along the [101] zone axis. This orientation was found to be the most frequent for the particles in the TEM samples and the results presented in Fig. 7 are representative, i.e., multiple particles were observed to have the same characteristics. There are a number of observations to note in the electron diffraction pattern. Firstly, the symmetry of the crystal structure of $Tb_2Cu_{0.83}Pd_{0.17}O_4$ differs slightly from *Pbca* because the (010) and (-101) reflections (along the [101] zone axis), forbidden in the *Pbca* space group, are observed by electron diffraction with weak intensities. Secondly, many superlattice reflections can clearly be seen along the orthorhombic *a* axis. This superstructure was measured to have a wavenumber $q \sim 0.47\ a$, which is incommensurate with the fundamental lattice. It is worthwhile to mention that the superstructure observed here is different, in terms of the $q$ value and the direction, from a few types of superstructures observed in other $R_2CuO_4$ compounds.[9] The corresponding real-space image is shown in Fig. 7b. This image was obtained from a thick area of the particle because it was observed that the superstructure is electron-beam sensitive in thin sample areas while stable in thick areas, which seems to be similar to the observations reported in other $R_2CuO_4$ compounds.[9] A Fast Fourier Transform (FFT) diffractogram is shown as an inset in Fig. 7b and is consistent with the electron diffraction pattern. Both the electron diffraction patterns and the images indicate that the incommensurate superstructure is unidirectional and long-range. It is not known whether the violations of the *Pbca* symmetry observed by electron diffraction are a consequence of this structural modulation or an additional structural distortion.

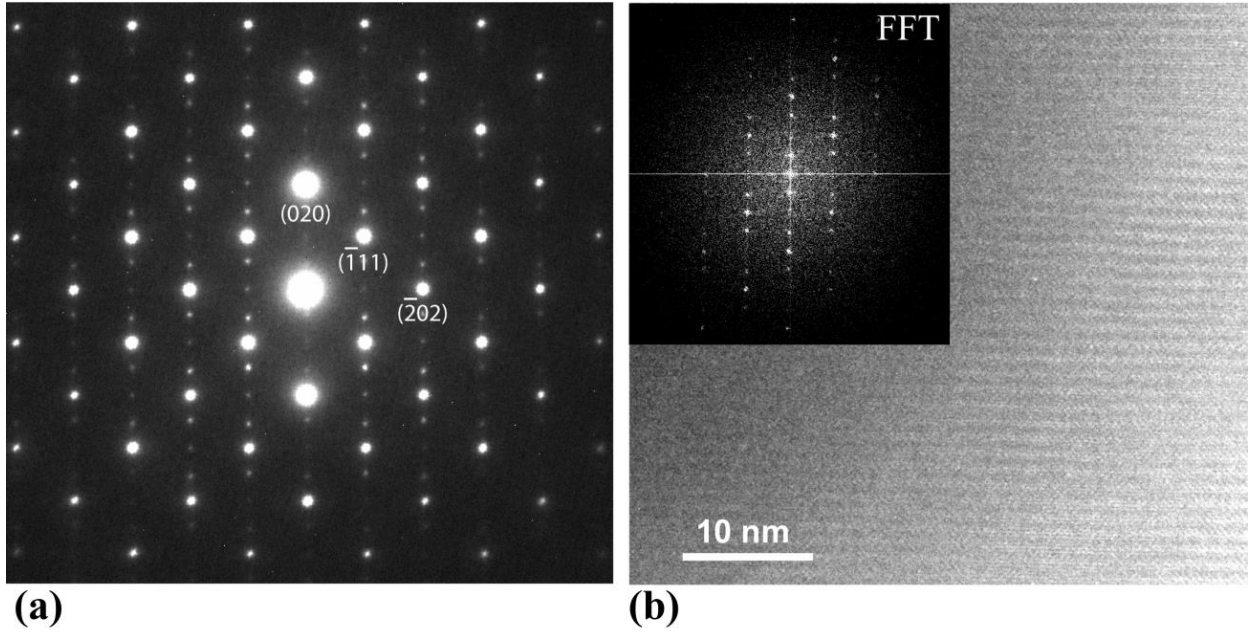

FIG. 7. (a) The electron diffraction pattern of the [101] zone axis of $Tb_2Cu_{0.8}Pd_{0.2}O_4$ at room temperature. In addition to the reflections such as (020), (-202), and (-111), which are allowed in the orthorhombic *Pbca* space group, reflections forbidden in that space group can also be identified including (010) and (-101) reflections with weak intensities. Additional extra spots, satellite peaks of the fundamental reflections, indicate that a superlattice exists along the *a* axis of the orthorhombic structure. (b) HRTEM images obtained from the same particle in (a), along the [101] zone axis, show superstructure fringes, another indication of the superlattice. The Fast Fourier Transform (FFT) diffractogram in the inset is consistent with the electron diffraction pattern.

Our temperature-dependent magnetic susceptibility data are shown in the main panel of Fig. 8 for $Tb_2Cu_{0.83}Pd_{0.17}O_4$ synthesized at 1050 °C in air. Measurements taken from 300 K to 1.7 K under a constant field of 1000 Oe were used to determine the magnetic susceptibility, M/H. To obtain the Curie-Weiss temperature ($\theta_{CW}$), the magnetic data were fit to the Curie-Weiss law $\chi = \frac{C}{T-\theta_{CW}}$, where $\chi$ is the magnetic susceptibility, C is the Curie constant, and T is the temperature. The effective moment was calculated using $\mu_{eff} \propto 2.83\sqrt{C}$. Inverse magnetic susceptibility data were fit in the range 150–300 K using the Curie-Weiss law, yielding a Curie constant of 23.8 and Weiss temperature of −18.6 K with a $\mu_{eff}$ of 13.8 $\mu_B$/F.U. Accordingly, $\mu_{eff}$ per Tb is 9.75 $\mu_B$/Tb, and is very close to the value expected for a free $Tb^{3+}$ ion. This negative Weiss temperature, which agrees with that of the reported Weiss temperature of the high-pressure synthesized $Tb_2CuO_4$ material, suggests that the sample has dominantly antiferromagnetic coupling between spins.[5] In fact, the ordering at 7.9 K with the magnetic susceptibility sharply decreasing is an indication of an antiferromagnetic transition. There is another, less pronounced feature at 95 K. Below this temperature, $\mu_{eff}$ is calculated as 15.5 $\mu_B$/F.U. The difference between the ZFC and FC data below 100 K shown in Fig. 8b, measured in an applied field of 50 Oe, suggests that the feature at about 95 K is due to the presence of canted antiferromagnetic ordering at this temperature, probably of the copper sublattice, as has been previously suggested for $Gd_2CuO_4$ and other $R_2CuO_4$ heavy rare earth variants.[2,13]

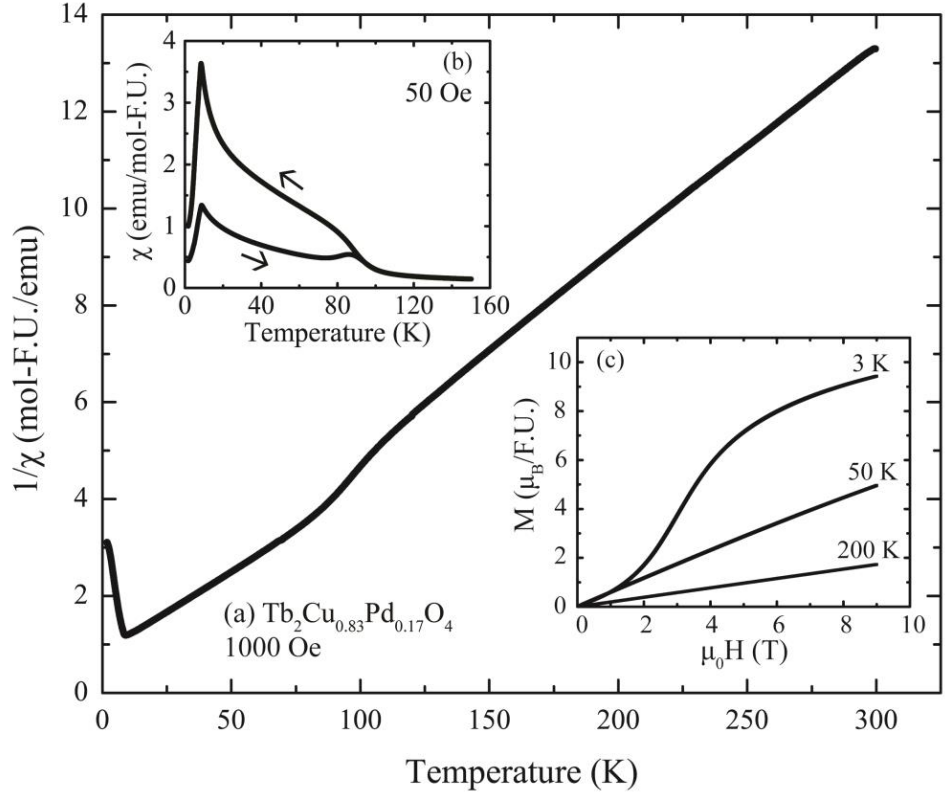

FIG. 8. Magnetic characterization of a powder sample of $Tb_2Cu_{0.83}Pd_{0.17}O_4$. (a) Inverse magnetic susceptibility in an applied field of 1000 Oe. The sample is antiferromagnetic with an ordering temperature of 7.9 K. For 150 K < T < 300 K the Curie-Weiss temperature $\theta_{CW} = -18.6$ K with a $\mu_{eff}$ of 13.8 $\mu_B$/F.U., while the $\mu_{eff}$ is 15.5 $\mu_B$/F.U. for 14 K < T < 75 K. (b) Magnetic susceptibility as a function of temperature for zero-field cooled and field-cooled material measured at 50 Oe. (c) Magnetization as a function of applied field at 3 K, 50 K, and 200 K. At 3 K the sample is antiferromagnetic, deviating from linear behavior at an applied field of approximately 1.1 T, while at 200 K the sample is paramagnetic. Detailed magnetic data taken at lower fields at 50 K (not shown) indicate the presence of a very small remnant magnetization and low coercive field at this temperature.

Field-dependent magnetization data for Tb$_2$Cu$_{0.83}$Pd$_{0.17}$O$_4$ taken at 3, 50, and 200 K are shown in Fig. 8c. At 3 K, the sample is antiferromagnetic (attributed to the rare earth spins - the copper spins are apparently ferrimagnetically aligned below 95 K, see below), with a broad metamagnetic transition starting at an applied field of approximately 1.1 T. The other two temperatures show linear increases in magnetization, characteristic of paramagnetism. This is consistent with the temperature dependent magnetization data. Moreover, these data are similar to those reported for other $R_2$CuO$_4$ T'-type materials ($R$ = Dy, Er, Ho, and Tm).[3] A plot of magnetization vs. applied field at 50 K (data not shown) indicates the presence of a very narrow hysteresis loop with a remanent magnetization of 54.3 emu/mol and a coercive field of 153.0 Oe, consistent with the presence of canted antiferromagnetic ordering of the Cu sublattice below 95 K.

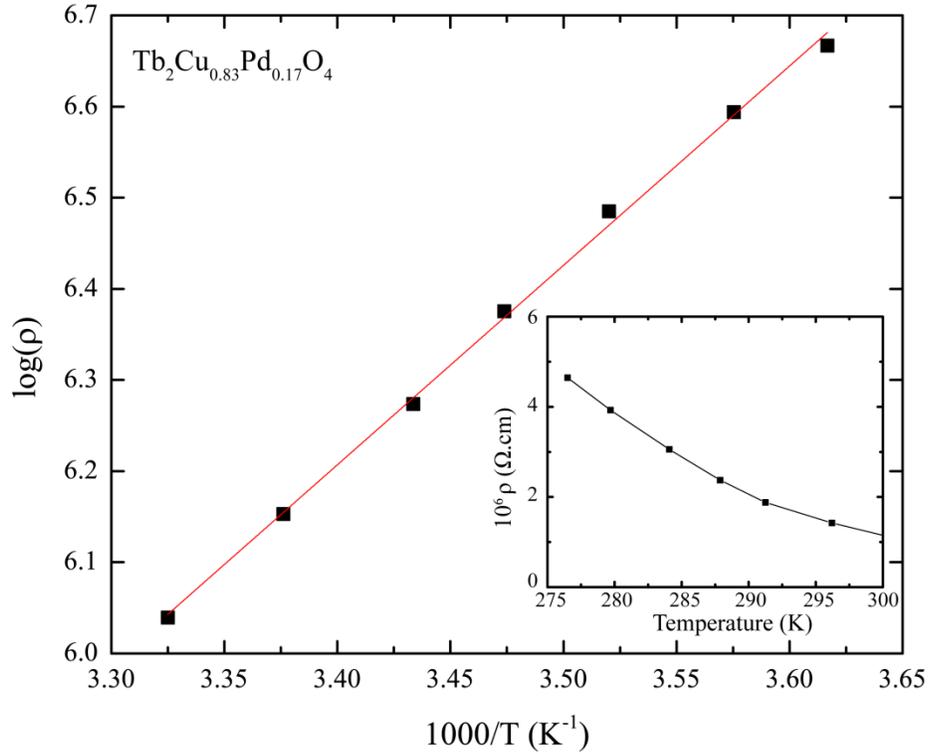

FIG. 9. Temperature dependent resistivity of a sintered polycrystalline pellet of Tb$_2$Cu$_{0.83}$Pd$_{0.17}$O$_4$ from 278–298 K. (Main) The resistivity activation energy calculated from the slope of log($\rho$) vs. 1/T is 0.43 eV. (Inset) The sample displays non-metallic behavior, with the resistivity exponentially increasing with decreasing temperature.

Finally, Fig. 9 shows the log $\rho$ vs. 1/T resistivity data for Tb$_2$Cu$_{0.83}$Pd$_{0.17}$O$_4$, with $\rho$ vs. T shown in the inset. The highly resistive nature of the sample limited the temperature range for measurement, as the resistance exceeded $5 \times 10^6$ $\Omega$ by 275 K. The steadily increasing resistivity with decreasing temperature is characteristic of a semiconducting material. The activation energy ($E_a$) of the thermally activated charge carriers was determined using the activation law $\rho \propto e^{(E_a/k_B T)}$, where $k_B$ is the Boltzmann constant. $E_a$ was calculated to be 0.43 eV from the slope of the log ρ vs. 1/T plot (Fig. 9). The Seebeck coefficient hot probe test characterized the sample as being *n*-type.

**Conclusion**

We have shown that the T' type Tb-based 214 cuprate can be synthesized at atmospheric pressure by substituting 20% of the Cu with Pd, consistent with our hypothesis that $Pd^{2+}$ should be a stabilizing ion for the T' structure type. To the best of our precision, the optimal nominal composition needed to prepare the best bulk material under our conditions is $Tb_{1.96}Cu_{0.8}Pd_{0.2}O_4$, although the refinement of the crystal structure of the T' material formed found the composition to be slightly different, at $Tb_2Cu_{0.83}Pd_{0.17}O_4$. Through a Rietveld refinement, the stabilized phase was determined to have a *Pbca* T'-type structure with $a = 5.5117(1)$ Å, $b = 5.5088(1)$ Å, $c = 11.8818(1)$ Å, with the larger orthorhombic cell primarily arising from the rotation of the $Cu/PdO_4$ squares. An incommensurate superlattice that approximately doubles the *a* axis was observed by electron diffraction and imaging, but its detailed interpretation is beyond the scope of the current study. Temperature-dependent magnetic susceptibility measurements show that antiferromagnetic ordering occurs at 7.9 K, which we speculate arises from an ordering of the Tb moments; the origin of the feature in the magnetic susceptibility at 95 K appears to be canted antiferromagnetic ordering of the Cu moments. Resistivity and Seebeck measurements indicate that the sample is an *n*-type semiconductor. Although our hypothesis that Pd substitution for Cu should stabilize $R_2CuO_4$ T' type 214 structures turned out to be correct for $R$ = Tb, which is directly at the border between high pressure and ambient pressure synthetic stability, we did not find stabilization by partial Pd/Cu substitution to be generally applicable for rare earths smaller than Tb.

## Acknowledgements

This material is based upon work supported by the National Science Foundation (Platform for the Accelerated Realization, Analysis, and Discovery of Interface Materials (PARADIM) under Cooperative Agreement No. DMR-1539918. The use of the Advanced Photon Source at Argonne National Laboratory was supported by the U. S. Department of Energy, Office of Science, Office of Basic Energy Sciences, under Contract No. DE-AC02-06CH11357. The work at Brookhaven was sponsored by the U.S. DOE BES, by the Materials Sciences and Engineering Division under Contract DE-SC0012704, and supported by the resources of the Center for Functional Nanomaterials at Brookhaven National Laboratory, which is a U.S. DOE Office of Science Facility.

# References

(1) Von Hk. Müller-Buschbaum; W. Wollschläger: Zur Kristallstruktur von $Nd_2CuO_4$. *Z. Anorg. Allg. Chem.* **414**, 76 (1975).

(2) X. Obradors; P. Visani; M. A. de la Torre; M. B. Maple; M. Tovar; F. Perez; P. Bordet; J. Chenavas; D. Chateigner: Rare-Earth Magnetic Ordering in the $R_2CuO_4$ Cuprates ($R$ = Tb, Dy, Ho, Er and Tm). *Physica C*, **213**, 81 (1993).

(3) H. Okada; M. Takano; Y. Takeda: Magnetic properties of $Nd_2CuO_4$-type $R_2CuO_4$ ($R$ = Y, Dy, Ho, Er, Tm) synthesized under high pressure: Weak ferromagnetism of $Y_2CuO_4$. *Physical Review B*, **42** (10), 6813 (1990).

(4) J. D. Thompson; S. W. Cheong; S. E. Brown; Z. Fisk; S. B. Oseroff; M. Tovar; D. C. Vier; S. Schultz: Magnetic Properties of $Gd_2CuO_4$ Crystals. *Physical Review B*, **39** (10), 6660 (1989).

(5) M. Tovar; X. Obradors; F. Pérez; S. B. Oseroff; R. J. Duro; J. Rivas; D. Chateigner; P. Bordet; J. Chenavas: Weak ferromagnetism and spin-glass-like behavior in $Tb_2CuO_4$. *Journal of Applied Physics*, **70** (10), 6095 (1991).

(6) Y. Tokura; H. Takagi; S. Uchida: A Superconducting Copper Oxide Compound with Electrons as the Charge Carriers. *Nature*, **337**, 345 (1989).

(7) M. B. Maple: Electron-Doped High $T_C$ Superconductors. *MRS Bulletin*, **15** (6), 60 (1990).

(8) H. Okada; M. Takano, Y. Takeda: Synthesis of $Nd_2CuO_4$-type $R_2CuO_4$ ($R$ = Y, Dy, Ho, Er, Tm) under High Pressure. *Physica C,* **166**, 111 (1990).

(9) P. Bordet; J. J. Capponi; C. Chaillout; D. Chateigner; J. Chenavas; T. Fournier; J. L. Hodeau; M. Marezio; M. Perroux; G. Thomas; A. Varela: High pressure synthesis and structural study of $R_2CuO_4$ compounds with R = Y, Tb, Dy, Ho, Er, Tm. *Physica C*, **193**, 178 (1992).

(10) H. M. Luo; Y. Y. Hsu; B. N. Lin; Y. P. Chi; T. J. Lee; H. C. Ku: Correlation between weak ferromagnetism and crystal symmetry in $Gd_2CuO_4$-type cuprates. *Physical Review B*, **60** (18), 119 (1999).

(11) M. Braden; W. Paulus; A. Cousson; P. Vigoureux; G. Heger; A. Goukassov; P. Bourges; D. Petitgrand: Structure analysis of $Gd_2CuO_4$: A new modification of the *T'* phase. *Europhy. Lett.,* **25** (8), 625 (1994).

(12) I. P. Makarova; V. I. Simonov; M. K. Blomberg; M. J. Merisalo: X-Ray diffraction study of $Nd_2CuO_4$ single crystals at 20 K. *Acta Crystallographica Section B*, **52**, 93 (1996).

(13) A. Rouco; X. Obradors; M. Tovar; P. Bordet; D. Chateigner; J. Chenavas: Magnetic-Field-Induced Weak Ferromagnetic Order in $Y_2CuO_4$. *Europhys. Lett.*, **20** (7), 651 (1992).